\documentclass{article}

\usepackage{amsmath,amssymb,graphicx}

\begin{document}

{\flushright {\tt AEI-2009-117\\ULB-TH/09-41}\\[20mm]}

\begin{center}
{\Large  \bf Cosmological Quantum Billiards}~\footnote{Based on a talk
   given by H.~Nicolai at the Conference ``Foundations of space and time:
   reflections on quantum gravity'' in honor of George Ellis' 70th
   birthday, Stellenbosch, South Africa, 10-14 August 2009.} \\[10mm]

Axel Kleinschmidt${}^*$ and Hermann Nicolai${}^\dagger$\\[7mm]
${}^*$Physique Th\'eorique et Math\'ematique \& International Solvay Institutes, 
Universit\'e Libre de Bruxelles, Boulevard du Triomphe, ULB-CP231,\\ BE-1050 Bruxelles, Belgium\\[5mm]

${}^\dagger$Max-Planck-Institut f\"ur Gravitationsphysik, Albert-Einstein-Institut, \\
Am M\"uhlenberg 1, DE-14476 Golm, Germany\\[10mm]

\begin{tabular}{p{105mm}}
\noindent {\footnotesize
The mini-superspace quantization of $D=11$ supergravity is 
equivalent to the quantization of  a $E_{10}/K(E_{10})$ coset
space sigma model, when the latter is restricted to the $E_{10}$ Cartan 
subalgebra. As a consequence, the  wavefunctions solving the relevant
mini-superspace Wheeler-DeWitt equation involve automorphic (Maass wave) 
forms under the modular group $W^+(E_{10}) \cong PSL_2(\mathtt{O})$. 
Using Dirichlet boundary conditions on the billiard domain a 
general inequality for the Laplace eigenvalues of these automorphic 
forms is derived, entailing a wave function of  the universe that 
is generically complex and always tends to zero when approaching 
the initial singularity. The significance of these properties for 
the nature of singularities in quantum cosmology in comparison 
with other approaches is discussed. The present approach also offers
interesting new perspectives on some long standing issues in canonical
quantum gravity.}
\end{tabular}

\vspace{10mm}
\end{center}

\begin{section}{Introduction}
The present contribution is based on \cite{KKN}, and elaborates on
several issues and arguments that were not fully spelled out there.
In that work, a first step was taken towards quantization of the
one-dimensional `geodesic' $E_{10}/K(E_{10})$  coset model which had 
been proposed in \cite{Damour:2002cu} as a model of M-theory. Here, 
$E_{10}$ denotes the hyperbolic Kac--Moody group $E_{10}$ which is 
an infinite-dimensional extension of the exceptional Lie group $E_8$,
and plays a similarly distinguished role among the infinite-dimensional
Lie algebras as $E_8$ does among the finite-dimensional ones. The proposal 
of~\cite{Damour:2002cu} had its roots both in the appearance of so-called 
`hidden symmetries' of exceptional type in the dimensional reduction of 
maximal supergravity to lower dimensions \cite{CJ}, as well as in the 
celebrated analysis of Belinskii, Khalatnikov and Lifshitz (BKL)~\cite{BKL} 
of the gravitational field equations in the vicinity of a generic 
space-like (cosmological) singularity. According to the basic hypothesis 
underlying this analysis the causal decoupling of spatial points near 
the spacelike singularity\footnote{This is the same decoupling that later came 
  to be associated with the so-called `horizon problem' of inflationary 
  cosmology.} effectively leads to a dimensional reduction whereby 
the equations of motion become ultralocal in space, and the dynamics
should therefore be describable in terms of a (continuous) superposition 
of one-dimensional systems, one for each spatial point. More specifically, 
in this approximation the dynamics at each spatial can be described by 
a sequence of Kasner regimes, such that in the strict limit towards 
the singularity, the Kasner behavior is interspersed with hard 
reflections of the logarithms of the spatial scale factors off 
infinite potential walls~\cite{CWM,RS}. This generic behavior has 
been termed `cosmological billiards'. The geometry of the billiard 
table and the configuration of the walls (`cushions') of the billiard 
table are determined by the dimension and the matter content of 
the theory~\cite{CosmoBilliards,LivRev}. Likewise the occurrence 
or non-occurrence of chaotic oscillations near the singularity depends
on this configuration. In particular, for $D=11$ supergravity it was 
shown in \cite{Damour:2000hv} that the billiard domain is the 
fundamental Weyl chamber $\mathcal{C}$ of the `maximally extended'
hyperbolic Kac--Moody group $E_{10}$. The volume of this fundamental Weyl 
chamber is finite, implying chaotic behavior~\cite{Damour:2000hv,DHJN}.
The emergence of the hyperbolic Kac--Moody algebra $E_{10}$ in this 
context is also in line with its conjectured appearance in the dimensional
reduction of $D=11$ supergravity to one time dimension \cite{Julia}.

Ref.~\cite{Damour:2002cu} goes beyond the standard BKL analysis,
as well as the original conjecture \cite{Julia}, in that it establishes a 
correspondence at the classical level between a truncated gradient 
expansion of the $D=11$ supergravity equations of motion near the
spacelike singularity and an expansion in heights of roots of a one-dimensional
constrained `geodesic' $E_{10}/K(E_{10})$ coset space model. The cosmological 
billiards approximation then corresponds to the restriction of this 
coset model to the Cartan subalgebra of $E_{10}$. Going beyond this
billiard approximation involves bringing in spatial dependence, in such 
a way that a `small tension' expansion in spatial gradients {\em \`a la}
BKL gets converted into a level expansion of the $E_{10}$ Lie algebra. 
However, the correspondence between the field equations on the 
one hand, and the $E_{10}/K(E_{10})$ model on the other hand, codified in 
a `dictionary', has so far only been shown to work up to first order 
spatial gradients. The proper inclusion of higher order spatial gradients, 
and thus the emergence of a space-time field theory from a `pre-geometrical'
scheme, remains an outstanding problem, in spite of the fact that 
the $E_{10}$ Lie algebra contains all the `gradient representations' 
that would be needed for a Taylor expansion around a given spatial 
point~\cite{Damour:2002cu}.

Quantizing M-theory in the $E_{10}$ framework thus amounts to setting up
and solving a Wheeler-DeWitt equation for the full $E_{10}/K(E_{10})$ 
model, and imposing the subsidiary constraints corresponding to the 
canonical (diffeomorphism, Gauss,...) constraints of the usual 
canonical approach. As a very first step, ref.~\cite{KKN}
solves the quantum constraints for $D=11$ supergravity for the 
ten spatial scale factors, and for their fermionic `superpartners', 
in compliance with the supersymmetry constraint. The 
resulting {\em quantum cosmological billiard} is
a variant of the `minisuperspace'  quantization  of gravity pioneered in 
 \cite{Minisuperspace,Misner,Misner:1972js} and further developed in 
 many works, see in particular
\cite{Macias:1987ir,Graham:1990jd,Ivashchuk:1994fg,Moniz:1996pd,D'Eath:1996at,Csordas:1995kd,Kiefer}
and references therein.\footnote{We remind readers that the `super' in 
  `minisuperspace' refers to Wheeler's moduli space of 3-geometries
  and is a purely  bosonic concept. It therefore has no relation to the 
  notion of `supersymmetry' relating bosons and fermions, which is also
  discussed in this article.}  The essential new ingredient in the present 
construction is the {\em arithmetic structure} provided by $E_{10}$ 
and its Weyl group, whose relevance in the context of Einstein gravity 
was pointed out and explored in \cite{Forte:2008jr}; see also \cite{Pioline}
for an ansatz based on superconformal quantum mechanics, which has some
similarities with the present approach. We believe that
the key issues with the proposal of \cite{Damour:2002cu}, in particular 
the conjectured emergence of classical space-time out of a pre-geometrical 
quantum gravity phase, and the question how the `dictionary' of 
\cite{Damour:2002cu} can be extended to the full $E_{10}$ algebra, 
cannot be resolved within the framework of classical field equations, 
but will require quantization of the $E_{10}/K(E_{10})$ coset model. 
Equally important, the `resolution' of the cosmological singularity, 
a key issue of modern research in canonical quantum gravity, is expected
to involve quantum theory in an essential way. In addition, it will 
almost certainly require new concepts besides the transition to the 
quantum theory, with the question of what happens to classical space-time 
near the singularity as the core problem.

The main results reported in~\cite{KKN} and in the present contribution are
\begin{itemize}
\item The quantum cosmological billiard problem is well-posed and allows for a Hilbert space (with positive definite metric).
\item The solutions to the bosonic Wheeler-DeWitt (WDW) equation of $D=11$ 
      supergravity can be described as odd Maass wave forms of the 
      `modular group' $W^+(E_{10})\cong PSL(2,\mathtt{O})$.
\item With the appropriate Dirichlet boundary conditions all solutions of the WDW equation, 
      a.k.a. `wavefunctions of the universe', can be shown generally to 
      vanish rapidly near a space-like singularity while remaining complex 
      and oscillating.
\item The analysis can be extended to include the fermionic degrees of 
      freedom without changing the conclusions.
\end{itemize}
Possible extensions and the eventual significance of these results for
quantum cosmology are discussed at the end of this article.

\end{section}

\begin{section}{Minisuperspace quantization}
We first consider the bosonic variables. We proceed from the following metric 
ansatz, as appropriate for a cosmological billiard for a $(d+1)$-dimensional 
space-time 
\begin{eqnarray}
ds^2 = -N^2 dt^2 + \sum_{a=1}^{d} e^{-2\beta^a} \theta^a\otimes \theta^a\,,
\end{eqnarray}
where we keep the spatial dimension $d$ arbitrary (but always $d\geq 3$)
for the moment, and will specialize to $D=11$ supergravity (and
thus $d=10$) only later. $\theta^a \equiv {\cal{N}}^a{}_m dx^m$ is
a spatial frame in an Iwasawa decomposition of the metric, as explained
in \cite{CosmoBilliards}. Substituting the above ansatz into the 
Einstein action, one arrives at the kinetic term
\begin{eqnarray}
\mathcal{L}_{\text{kin}} = \frac12 n^{-1} 
\sum_{a,b=1}^{d} \dot{\beta}^a G_{ab} \dot{\beta}^b
\end{eqnarray}
in terms of the new lapse $n\equiv N/\sqrt{g}$ (the spatial volume is 
$\sqrt{g} =\exp [-\sum_a\beta^a]$) and the Lorentzian DeWitt metric
\begin{eqnarray}\label{DW}
\dot{\beta}^a G_{ab} \dot{\beta}^b \equiv 
\sum_{a=1}^{d} (\dot{\beta^a})^2 - 
\left(\sum_{a=1}^{d} \dot{\beta}^a\right)^2\,.
\end{eqnarray}
As is well known, the DeWitt metric, with the factor $(-1)$ in front of 
the second term, is distinguished by several uniqueness properties 
that are discussed in~\cite{GK}. Here, it 
will be essential that for $d=10$ this metric coincides with the 
restriction of the Cartan--Killing metric of $E_{10}$ to its Cartan 
subalgebra. It can now be shown~\cite{CosmoBilliards} that the remaining
contributions to the Hamiltonian constraint at a given spatial point 
can be combined into an `effective potential' of the generic form
\begin{eqnarray}
\label{veff}
V_{\text{eff}} = \sum_A c_A(Q,P,\partial\beta,\partial Q)
                 \exp\big(-2w_A(\beta)\big)
\end{eqnarray}
where $(Q,P)$ are the (canonical) variables corresponding to all degrees
of freedom other than the scale factors $\beta^a$, the $\partial$ stands 
for {\em spatial} gradients, and $w_A$ in the exponent are  linear forms, called {\em wall forms},
\begin{eqnarray}
w_A(\beta) \equiv G_{ab} w_A^a \beta^b
\end{eqnarray}
with the DeWitt metric $G_{ab}$ introduced above. In the limit
towards the singularity $\beta\rightarrow\infty$, the exponential walls
become `sharp', and the dynamics is dominated by a set of `nearest' walls.
These make up the `cushions' of a billiard table, and can be viewed 
as the result of `integrating out' the off-diagonal metric and the matter 
degrees of freedom, as explained in \cite{CosmoBilliards}. In the strict 
limit towards the singularity they are simply given by timelike hyperplanes 
in the forward lightcone in $\beta$-space, which are determined by
the linear equations $w_A(\beta) =0$. The spatial ultralocality of 
the BKL limit thus reduces the gravitational model to a classical 
mechanics system of a relativistic billiard ball described by the 
$\beta^a$ variables moving on straight null lines in the Lorentzian 
space with metric $G_{ab}$ until hitting a billiard table wall corresponding
to a hyperplane. The straight line segments of the billiard motion are 
the Kasner regimes, while the reflections are usually referred to as 
`Kasner bounces'. We repeat that there is one such system for each 
spatial point ${\bf x}$, and these systems are all decoupled.

The canonical bosonic variables of the billiard system are $\beta^a$ and 
their canonically conjugate momenta $\pi_a$, {\it viz.}
\begin{eqnarray}
\pi_a := \frac{\partial \mathcal{L}}{\partial \dot {\beta}^a}= G_{ab}\dot{\beta}^b  
\end{eqnarray}
where we set $n=1$ from now on.\footnote{With this choice of gauge,
  $t$ becomes a `Zeno-like' time coordinate, for which the singularity 
  is at $t=+\infty$. This time is related to physical (proper) time 
  $T$ by $t\sim-\log T$.} The Hamiltonian is 
\begin{eqnarray}
\mathcal{H}_0 =\frac12\pi_a G^{ab} \pi_b
\end{eqnarray} 
with the inverse DeWitt metric $G^{ab}$. The effective potential (\ref{veff}) has disappeared as we have taken the BKL limit.
Before quantisation, we perform the following change of variables  
by means of which the billiard motion is projected onto the unit 
hyperboloid in $\beta$-space \cite{CosmoBilliards}
\begin{eqnarray}\label{CT}
\beta^a = \rho \omega^a\,,\quad \omega^a G_{ab} \omega^b =-1\,,\quad 
\rho^2=-\beta^aG_{ab}\beta^b\,,
\end{eqnarray}
where $\rho$ is the `radial' direction in the future light-cone and 
$\omega^a=\omega^a(z)$ are expressible as functions of $d-1$ coordinates 
$z$ on the unit hyperboloid. The limit towards the singularity is 
$\rho\to\infty$ in these variables. The Wheeler-DeWitt operator 
thus takes the form
\begin{eqnarray}\label{WDWHam}
\mathcal{H}_0 \equiv G^{ab}\partial_a \partial_b = 
 -\rho^{1-d}\frac{\partial}{\partial\rho}\left(\rho^{d-1}
 \frac{\partial}{\partial\rho}\right) + \rho^{-2} \Delta_{\text{LB}}\,,
\end{eqnarray}
where $\Delta_{\text{LB}}$ is the Laplace--Beltrami operator on the 
$(d-1)$-dimensional unit hyperboloid. We emphasize that ordering 
ambiguities are entirely absent in this expression, as is manifest in
the $\beta^a$ variables in terms of which the WDW equation is just
the free Klein-Gordon equation. The same holds true for the variables
$(\rho,\omega^a(z))$ because the expression in the new coordinates
is unambiguously determined by the coordinate transformation (\ref{CT})
(as it would be in any other coordinate system).

The mini-superspace WDW equation therefore reads 
\begin{eqnarray}
\mathcal{H}_0  \Phi(\rho,z)=0
\end{eqnarray} 
for the `wavefunction of the universe' $\Phi(\rho,z)$. As usual 
(see e.g. \cite{Kiefer}) one can adopt $\rho$ as a time coordinate 
in the initially `timeless' WDW equation, with the standard 
(Klein--Gordon-like) invariant inner product 
\begin{eqnarray}\label{KGProduct}
(\Phi_1,\Phi_2) = i \int d\Sigma^a 
\Phi_1^* \stackrel{\leftrightarrow}{\partial_a} \Phi_2
\end{eqnarray}
where the integral is to be taken over a spacelike hypersurface inside
the forward lightcone in $\beta$-space. `Invariance' means that this scalar
product does not depend on the shape of the spacelike hypersurface, so
we can  for instance choose any of the unit hyperboloids $\rho=const$.

In order to construct solutions we separate variables by means of the ansatz
$\Phi(\rho,z)= R(\rho)F(z)$ \cite{Misner:1972js,Graham:1990jd}. 
For any eigenfunction $F (z)$ obeying
\begin{eqnarray}\label{Laplace}
-\Delta_{\text{LB}} F (z) = E F(z)
\end{eqnarray}
the associated  radial equation is solved by
\begin{eqnarray}\label{R}
R_\pm(\rho) = \rho^{-\frac{d-2}{2}} 
e^{ \pm i  \sqrt{E-\left(\frac{d-2}{2}\right)^2}\log\rho}\,.
\end{eqnarray}
Positive frequency waves emanating from the singularity correspond 
to $R_-(\rho)$ and have positive inner product (\ref{KGProduct}). It 
is important here that one can consistently restrict to positive norm 
wave functions: the potential which might scatter an initially positive 
norm state into a negative norm state, is here effectively replaced by 
a set of boundary conditions on the wavefunction, and hence there is 
no `Klein paradox'. As we will see the same feature continues to hold for 
the full $E_{10}$ WDW operator, in marked contrast to the standard 
WDW operator. 

To study the eigenvalues of the Laplace--Beltrami operator on the unit 
hyperboloid we use a generalized upper half plane model $z=(\vec{u},v)$ 
for the unit hyperboloid with coordinates $\vec{u}\in \mathbb{R}^{d-2}$ 
and $v\in \mathbb{R}_{>0}$. The relevant coordinate transformation is
obtained by first diagonalizing the DeWitt metric (\ref{DW}) in terms of
Minkowskian coordinates $\tilde\beta^a$, such that
\begin{eqnarray}
G_{ab}\beta^a\beta^b \equiv - \tilde\beta^+\tilde\beta^- + 
\sum_{j=1}^{d-2} \tilde\beta^j\tilde\beta^j \,,
\end{eqnarray}
where we have used the last two directions for forming light-cone coordinates.
Then the forward unit hyperboloid (with $\tilde\beta^\pm >0$)
is coordinatized by
\begin{eqnarray}\label{tomega}
\tilde\beta^+= \frac1{v} \;,\quad
\tilde\beta^- = v + \frac{\vec{u}^{\,2}}{v}\;,\quad
\tilde\beta^j = \frac{u^j}{v} \qquad (v>0)
\end{eqnarray}
The metric induced on the unit hyperboloid is easily calculated to
be the Poincar\'e metric on the generalized upper half plane
\begin{eqnarray}
ds^2 = \frac{dv^2 + d\vec{u}^{\, 2}}{v^2} \quad\Rightarrow\quad
d{\rm vol} (z) \equiv v^{1-d} \, dv d^{d-2}u
\end{eqnarray}
such that the Laplace-Beltrami operator becomes
\begin{eqnarray}\label{DLB}
\Delta_{\text{LB}} = v^{d-1} \partial_v\left(v^{3-d}\partial_v\right) 
+ v^2 \partial_{\vec{u}}^2\,.
\end{eqnarray}
For the spectral problem we must specify boundary conditions. For the
cosmological billiard, these are provided by infinite (`sharp') 
potential walls which encapsulate the effect of spatial inhomogeneities
and matter fields near the spacelike singularity, as explained above
(see \cite{CosmoBilliards,LivRev} for details). Following the original 
suggestion of \cite{Misner:1972js}, we are thus led to impose the vanishing 
of the wavefunction on the boundary of the fundamental domain specified 
by these walls. Accordingly, let $F(z)$ be any function on the hyperboloid 
satisfying (\ref{Laplace}) with {\em Dirichlet conditions}  at the 
boundaries of this domain (in contrast to \cite{Graham:1990jd,Forte:2008jr}, 
where Neumann boundary conditions are assumed). A direct generalization 
of the arguments on page~28 of Ref.~\cite{Iwaniec} gives
\begin{eqnarray}
-(\Delta_{\text{LB}}F , F ) 
\geq \int dv\, d^{d-2}u\, v^{3-d} (\partial_v F)^2\,
\end{eqnarray}
with (\ref{Laplace}) and (\ref{DLB}). Considering also
\begin{eqnarray}
(F,F) &\equiv & \int d{\rm vol}(z) F^2(z) = 
\int dv\,d^{d-2}u\, v^{1-d} F^2\nonumber\\
& =& \frac2{d-2} \int dv\, d^{d-2}u\,v^{2-d} F \partial_v F\,,
\end{eqnarray}
the use of  the Cauchy--Schwarz inequality entails
\begin{eqnarray}\label{eigenvalues}
E \geq \left(\frac{d-2}{2}\right)^2\,.
\end{eqnarray}
From the explicit solution (\ref{R}) we thus conclude that $R_\pm(\rho)\to 0$ 
when $\rho\to\infty$, and therefore {\em the full wavefunction and all its 
$\rho$ derivatives tend to zero near the singularity}. Evidently, this 
result hinges on the peculiar form of the differential operator in the
$(\rho,z)$ variables in (\ref{WDWHam}), which itself is uniquely
determined by the form of the operator in $\beta$-coordinates. It
would not be valid if we were allowed to move around the $\rho$ factors
in the differential operator of (\ref{WDWHam}).

While the wave function would also vanish for Neumann boundary conditions 
(for which $E\geq 0$) with the given ordering, the inequality 
(\ref{eigenvalues}) furthermore ensures that the full wavefunction 
is generically complex and oscillating. Let us point out here that 
this result may be of relevance to a long standing issue in canonical 
gravity, namely the question why and how the {\em real} WDW equation
should give rise to {\em complex} wave functions~\cite{Isham,Barbour1,Barbour}.
As explained there, the complexity of the wave function is intimately 
linked to the emergence of a {\em directed} time in canonical gravity. More 
specifically, admitting only positive norm wave functions corresponds 
to choosing an `arrow of time' (real wave functions have vanishing norm 
with the product (\ref{KGProduct}), and would thus not select a time
direction). Let us repeat that restricting to positive norm states 
would be inconsistent for the standard WDW equation with a potential 
even in mini-superspace quantisation. Here, the potential has effectively 
disappeared in the BKL limit, leaving its trace only via the boundary
conditions, so the restriction is consistent.
\end{section}

\begin{section}{Automorphy and the $E_{10}$ Weyl group}

Whereas the discussion above was valid for gravity in any space-time of dimension $d+1$ we now focus on maximal supergravity in eleven space-time dimensions. For the bosonic sector of maximal supergravity, the wavefunctions can
be further analyzed by exploiting the underlying symmetry encoded in the 
Weyl group $W(E_{10})$ and its arithmetic properties, and in particular 
the new links between hyperbolic Weyl groups and generalized modular groups
uncovered in \cite{Feingold:2008ih}. The Weyl reflections that the classical 
particle is subjected to when colliding with one of the walls are norm 
preserving, and therefore the reflections can be projected to any hyperboloid 
of constant $\rho$, inducing a non-linear action on the co-ordinates 
$z$ (given in (\ref{wz}) below for the fundamental reflections).
For physical amplitudes to be invariant under the Weyl group, 
the full wavefunction must transform as follows 
\begin{eqnarray}\label{wPhi}
\Phi(\beta) = \pm \Phi (w_I(\beta)) \quad\Leftrightarrow\quad
\Phi(\rho,z) = \pm \Phi\big(\rho,w_I (z)\big)
\end{eqnarray}
for the ten generating fundamental reflections $w_I$ of $W(E_{10})$, 
labeled by $I=-1,0,1,\ldots,8$. Restricting the wavefunction to the 
fundamental Weyl chamber, one easily checks that
the plus sign in (\ref{wPhi}) corresponds to Neumann boundary conditions, 
and the minus sign to Dirichlet conditions (which we adopt here).
From (\ref{wPhi}) it follows that $\Phi(\rho,z)$ is invariant under 
{\em even} Weyl transformations $s\in W^+(E_{10})$ irrespective of the 
chosen boundary conditions.

Choosing coordinates as in (\ref{tomega}) the relevant variables
now live in a nine-dimensional `octonionic upper half plane' with 
coordinate
\begin{eqnarray}
z=u+\text{i} v \quad,  \qquad\quad   u\equiv\vec{u}\in\mathbb{O} 
\end{eqnarray}
where $\mathbb{O}$ is the non-commutative and non-associative algebra
of {\em octonions}, while $v$ is still real and positive. Next, we 
recall~\cite{Coxeter,Conway,Feingold:2008ih} that the 240 roots of $E_8$
can be represented by unit octonions; more precisely, these are the 
240 {\em units} in the non-commutative and non-associative ring of 
{\em integral octonions} $\mathtt{O}$ called {\em `octavians'}, 
see \cite{Conway}. Denoting by $\varepsilon_j$ (for $j=1,...,8$) the 
eight simple roots and by $\theta$ the highest root of $E_8$, respectively,  
expressed as unit octonions, we arrive at the following 
{\em modular realization} of the $E_{10}$ Weyl transformations 
on the nine-dimensional unit hyperboloid: the ten fundamental 
reflections of $W(E_{10})$ act as 
\begin{eqnarray}\label{wz}
w_{-1}(z)= \frac1{\bar{z}} \;,\quad
w_0 (z) = - \theta\bar{z}\theta + \theta \;,\quad
w_j (z) = -\varepsilon_j\bar{z}\varepsilon_j
\end{eqnarray}
where $\bar{z}:= \bar{u}-\text{i} v$, with $\text{i} u=\bar{u}\text{i}$ 
in accordance 
with Cayley--Dickson doubling \cite{Conway}.\footnote{We recall that the 
 Dickson doubling of a normed algebra $\mathbb{A}$ with conjugation 
 associates to doubled elements $a+\text{i}b, c+\text{i}d\in\mathbb{A}
 +\text{i}\mathbb{A}$ the product (see \cite{Conway})
 \begin{equation*}
 (a+\text{i}b) (c+\text{i}d) = (ac -d\bar{b}) +\text{i}(cb+\bar{a}d)\,,
 \end{equation*}
 conjugation being defined by $\overline{a+\text{i}b}=\bar{a}-\text{i}b$.
 The Hurwitz theorem states that, starting from the real numbers 
 $\mathbb{R}$, this process generates the division algebras 
 $\mathbb{R},\mathbb{C},\mathbb{H},\mathbb{O}$ of the real, complex, 
 quaternionic and octonionic numbers, respectively. Further application 
 generates algebras with zero divisors.} Observe that, despite the 
non-associativity of the octonions, there is no need to put parentheses 
in (\ref{wz}) by virtue of the alternativity of the octonions. 

In the present context, the formulas (\ref{wz}) represent the most 
general (and most sophisticated!) modular realization of a Weyl group, 
but there are corresponding versions for the other division algebras, 
with the quaternions $\mathbb{H}$ for $d=6$, and the complex numbers 
$\mathbb{C}$ for $d=4$ (with corresponding `integers', see 
\cite{Feingold:2008ih} for details). 
The simplest case is $\mathbb{A} =\mathbb{R}$ which corresponds 
to pure gravity in four spacetime dimensions ($d=3$). 
In this case $u\in\mathbb{R}$, and the formulas (\ref{wz}) reduce 
to the ones familiar from complex analysis, namely 
\begin{eqnarray}
z\mapsto \frac1{\bar{z}} \; ,\quad z\mapsto -\bar{z}+1 \; ,\quad
z\mapsto -\bar{z} \;, 
\end{eqnarray}
generating the group $PGL_2(\mathbb{Z})$. For {\em even} Weyl transformations, 
we re-obtain the standard modular group $PSL_2(\mathbb{Z})$ generated by 
\begin{eqnarray}
S(z)\equiv (w_{-1}w_1)(z)= -1/z \; , \qquad
T(z)\equiv (w_{0} w_1)(z) = z+1 \, . 
\end{eqnarray}
Consequently, for pure gravity in four space-time dimensions, 
the relevant eigenfunctions of the mini-superspace WDW operator 
are automorphic forms with respect to the standard modular 
group $PSL_2(\mathbb{Z})$, as already pointed
out in \cite{Forte:2008jr}. 

For the maximally supersymmetric theory, on the other hand, the even 
Weyl group $W^+(E_{10})$ is isomorphic to the `modular group' 
$PSL_2(\mathtt{O})$ over the octavians, where $PSL_2(\mathtt{O})$ is 
{\em defined} by iterating the action of (\ref{wz}) an even number 
of times \cite{Feingold:2008ih}. Accordingly, for maximal supergravity 
the bosonic wavefunctions $\Phi(\rho,z)$ are {\em odd Maass wave forms} 
for $PSL_2(\mathtt{O})$, that is, invariant eigenfunctions of the 
Laplace--Beltrami operator transforming with a minus sign in (\ref{wPhi}) 
under the extension $W(E_{10})$ of $PSL_2(\mathtt{O})$.
Properly understanding the `modular group' $PSL_2(\mathtt{O})$ and the
associated modular functions remains an outstanding mathematical 
challenge, see \cite{Eie:1992} for an introduction (and \cite{Iwaniec} 
for the $PSL_2(\mathbb{Z})$ theory). For the groups $PSL_2(\mathbb{Z})$ and 
$PSL_2(\mathbb{Z}[i])$ the (purely discrete) spectra of odd Maass wave forms 
have been investigated numerically in 
\cite{Steil,Hejhal,Bogomolny,Then2,Aurich:2004ik}. 

One important feature of (\ref{wz}) should be emphasized: supplementing 
the seven imaginary units of $\mathbb{O}$ by another imaginary unit 
$\text{i}$, it would appear that we have to enlarge the octonions to
{\em sedenions}, a system of hypercomplex numbers with 15 
imaginary units, which by Hurwitz' theorem is no longer a division 
algebra (that is, has zero divisors). Remarkably, however, the formulas 
(\ref{wz}) are such that with the iterated action of (\ref{wz}) we never 
need to introduce any further imaginary units beyond $\text{i}$ and the
seven octonionic ones. In other words, the transformations (\ref{wz}) 
do not move $z$ out of the 9-dimensional generalized upper half plane. 
In particular, the doubling rule ensures that 
$\bar{z}z = v^2 + |u|^2 \in {\mathbb{R}}_+$ so that the inverse $1/\bar{z}$ 
also stays in this plane. 

By modular invariance, the wavefunctions can be restricted 
to the fundamental domain  of the action of $W(E_{10})$ and, conversely,
their modular property defines them on the whole hyperboloid. 
The Klein--Gordon inner product (\ref{KGProduct}) must likewise 
be restricted to the fundamental chamber 
\begin{eqnarray}
(\Phi_1, \Phi_2) = i \int_{\mathcal{F}} d {\rm vol}(z)  
\rho^{d-1} \Phi_1^* \stackrel{\leftrightarrow}{\partial_\rho} \Phi_2\,.
\end{eqnarray}
where $\mathcal{F}$ is the intersection of $\mathcal{C}$ with the unit 
hyperboloid; accordingly, the `cushions' of the billiard are obtained
by intersecting the hyperplanes $w_A(\beta)=0$ with the unit hyperboloid,
such that for pure Einstein gravity ($d=3$) one ends up with the projected billiard table
shown in Figure~\ref{billiardfig}. The restriction of the scalar product to the fundamental
domain is necessary, as the integral over the whole hyperboloid would 
be infinite for functions obeying (\ref{wPhi}), and the product would be 
ill-defined. This infinity is analogous to the one that arises in the 
calculation of the one-loop amplitude in string theory, and there as well, 
the product is rendered finite upon `division' by the modular group 
$PSL_2(\mathbb{Z})$. We thus have at hand an analog of this mechanism in 
canonical quantum gravity. We note that modular invariance is a distinctive 
feature of string theory not shared by the quantum field theory of 
pointlike particles and arguably the `real' reason behind the conjectured 
finiteness of string theory. 
\end{section}

\begin{section}{Classical and quantum chaos}

\begin{figure}[t]
\centering
\includegraphics[scale=.2]{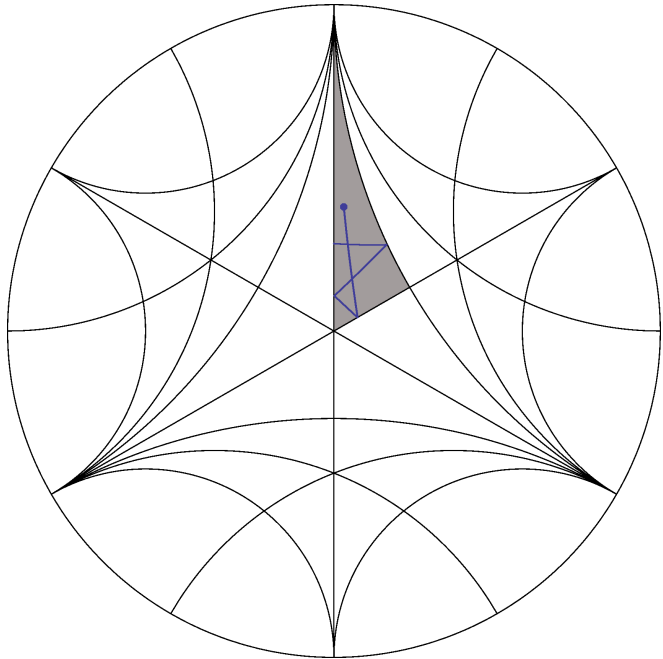}
\caption{\label{billiardfig} \em The projected billiard table of standard general relativity (in grey) on two-dimensional hyperbolic space. There are infinitely many equivalent tables related by Weyl reflections.}
\end{figure}

The fundamental region defined by the billiard walls as a subset of
the unit hyperboloid (a hyperbolic space of constant negative curvature) 
has finite volume 
\begin{eqnarray}
{\rm Vol} (\mathcal{F}) = \int_{\mathcal{F}} d {\rm vol}(z) < \infty
\end{eqnarray} 
in spite 
of the fact that the domain $\mathcal{F}$ extends to infinity (has 
a `cusp'). The finiteness of the fundamental domain is a consequence
of the hyperbolicity of the Kac--Moody algeba $E_{10}$ (whereas
fundamental domains for non-hyperbolic indefinite Kac--Moody algebras
have infinite volume). Such finite volume billiards have been known
and studied for a long time, as they are known to exhibit classical 
chaos. They are also the prototypical examples for studying the transition
from classical to quantum chaos, and especially the question of
how the presence of classical chaos is reflected in the spectra 
of the corresponding (unitary) Hamiltonian operators, see for
example~\cite{Berry:1977,Berry:1981,Gutzwiller,SieberSteiner}.
One of the remarkable results of these investigations is that there is a 
qualitative difference between the wave functions of classically 
ergodic and classically periodic orbits: The latter have very drastic 
(density) fluctuations whereas the former appear more like randomized 
Gaussians~\cite{Berry:1977} and can be called quantum ergodic. Another 
notable feature is the appearance of so-called `scars' as remnants of 
classically periodic orbits~\cite{Heller,Gutzwiller}.\footnote{See 
  \cite{Forte:2008jr} for a possible physical interpretation of 
  these `scars' in quantum cosmology.}
These are regions of (relative) high probability in position space 
which appear related to the positions of classically periodic orbits. 

However, there are two main differences between these studies and the 
cosmological (quantum) billiards considered here, {\it viz.}
\begin{itemize} 
\item The cosmological billiard is {\em relativistic}, that is, the 
      classical evolution follows a Klein Gordon-like equation, instead
      of a non-relativistic Schr\"odinger equation (but see \cite{Berry1}
      for a discussion of relativistic neutrino billiards).
\item In the $\beta$-space description, the walls of the billiard move
      away from one another, and one would thus have to solve the equation
      with {\em time-dependent} boundary conditions in the $\beta$
      variables. By contrast, the projection (\ref{CT}) allows to 
      reformulate the problem with static boundary conditions, at the 
      expense of modifying the $\rho$-dependent part of the relativistic 
      Hamiltonian $-\partial_\rho^2$ to the right hand side of 
      (\ref{WDWHam}). For pure Einstein gravity in four dimensions, 
      the resulting fixed walls billiard system is displayed 
      in figure~\ref{billiardfig}.
\end{itemize}

Within this new setting it would be of much interest to study the 
fate of a generic wavepacket in our cosmological billiard system.
The expectation is that the quantum theory `washes out' the classical 
chaos in the sense that any initially localized wavepacket will 
eventually disperse when approaching the singularity, such that the
asymptotic (for $\rho\rightarrow\infty$) wave function will be spread
evenly over $\mathcal{F}$. For the non-relativistic case some 
studies of the evolution of wavepackets can be found 
in~\cite{DavisHeller}, where the focus was on non-generic classically 
periodic configurations. 

The particular case of interest in M-theory cosmology possesses  a 
nine-dimensional fundamental domain and has apparently not been considered 
in the literature so far. The chaotic quantum billiard being merely 
the quantum theory of a {\em finite-dimensional} subsystem, corresponding
to the Cartan subalgebra of an infinite-dimensional Kac--Moody algebra, 
such a study would however represent only a first step towards the
quantization of the full system, as already mentioned. There will thus 
arise many new issues, such as for example the link between a formally 
integrable system in infinitely many variables, and the chaoticity of 
a finite dimensional system obtained from it by projection to finitely
many variables (some comments on this issue can be found in 
\cite{CosmoBilliards}).

\end{section}

\begin{section}{Supersymmetry}
The quantum billiard analysis can be extended to maximal {\em super}gravity, 
with $d=10$, by supplementing the bosonic degrees of freedom with
a vector-spinor, the gravitino. In the $E_{10}$ approach, the latter 
corresponds to a spinorial representation of the `R-symmetry' group
$K(E_{10})$ in terms of which the Rarita--Schwinger equation of
$D=11$ supergravity (with the usual truncations) can 
be re-written as a $K(E_{10})$ covariant `Dirac equation'
\cite{Damour:2005zs,de Buyl:2005mt,Damour:2006xu}. When restricting 
to the diagonal metric degrees of freedom, the 
gravitino $\psi_\mu$ of $D=11$ supergravity performs a separate 
classical\footnote{In the sense that the gravitino is treated as
  a classical variable, not as an operator.}
fermionic billiard motion~\cite{Damour:2009zc}. This 
is most easily expressed in a supersymmetry gauge $\psi_t =\Gamma_t 
\Gamma^a\psi_a$ \cite{Damour:2005zs} and in the 
variables \cite{Damour:2009zc} (with $\Gamma_*=\Gamma^1\cdots\Gamma^{10}$)
\begin{eqnarray}\label{BilliardFermions}
\varphi^a = g^{1/4}  \Gamma_* \Gamma^a \psi^a   
\quad\text{(no sum on $a=1,\ldots,10$)}\,.
\end{eqnarray}
(recall that $g\equiv \exp(- 2\sum\beta^a)$). Using (\ref{BilliardFermions}) 
in conjunction with Eqn.~(6.3) of \cite{Damour:2006xu} the Dirac brackets 
between two gravitino variables become 
\begin{eqnarray}\label{DB1}
\big\{ \varphi^a_\alpha\,,\,\varphi^b_\beta\big\}_{\rm D.B.} 
= -2i G^{ab} \delta_{\alpha\beta}
\end{eqnarray}
where we have written out the $32$ real spinor components using the 
indices $\alpha, \beta$. We stress that it is precisely the inverse DeWitt 
metric $G^{ab}$, see (\ref{DW}), which appears in this equation!

The fermionic and bosonic variables are linked by the supersymmetry constraint
\begin{eqnarray}\label{susycon}
\mathcal{S}_\alpha \equiv \sum_{a,b=1}^{10} \dot{\beta}^a G_{ab} 
\varphi^b_\alpha = \sum_{a=1}^{10} \pi_a \varphi^a_\alpha =0\,.
\end{eqnarray}
The supersymmetry constraint implies the Hamiltonian constraint 
$\mathcal{H}_0 =0$ by closure of the algebra
\begin{eqnarray}
\frac14\left\{ \mathcal{S}_\alpha , \mathcal{S}_\beta \right\}_{\rm D.B.} 
 = -i \delta_{\alpha\beta} \mathcal{H}_0 \,.
\end{eqnarray}
In order to quantize this system we rewrite the 320 real gravitino 
components $\varphi^a_\alpha$ in terms of 160 complex ones according to
\begin{eqnarray}
\tilde\varphi_A^a := \varphi_A^a + i \varphi_{A+16}^a
\end{eqnarray}
for $A,B,..= 1,\dots, 16$, and replace the Dirac brackets (\ref{DB1}) 
by canonical anticommutators 
\begin{eqnarray}
\big\{ \tilde\varphi_A^a \,,\, (\tilde\varphi^\dagger)_B^b \big\} =
 2 G^{ab} \delta_{AB}\; , \quad
\big\{ \tilde\varphi_A^a \,,\, \tilde\varphi_B^b \big\} = 
\big\{ (\tilde\varphi^\dagger)_A^a \,,\, 
(\tilde\varphi^\dagger)_B^b \big\} = 0
\end{eqnarray}
to obtain a fermionic Fock space of dimension $2^{160}$ by application
of the fermionic creation operators $(\tilde\varphi^a_A)^\dagger$ to the 
vacuum $|\Omega\rangle$ defined by
\begin{eqnarray}\label{vacuum}
\tilde\varphi^a_A\, |\Omega\rangle = 0
\end{eqnarray}
For the supersymmetry constraint this amounts to the redefinition 
$\tilde{\mathcal{S}}_A = \mathcal{S}_A+ i \mathcal{S}_{A+16}$. Because
(\ref{vacuum}) implies $\tilde{\mathcal{S}}_A|\Omega\rangle =0$, the
quantum supersymmetry constraint is then solved by
\begin{eqnarray}\label{formalsol}
|\Psi\rangle = \prod_{A=1}^{16}\tilde{\mathcal{S}}^\dagger_A 
          \Big( \Phi(\rho,z) |\Omega\rangle \Big) \,,
\end{eqnarray}
if and only if the function $\Phi(\rho,z)$ is a solution of the 
WDW equation $\mathcal{H}_0 \Phi=0$.
While this solution is close to the `bottom of the 
Dirac sea', there is an analogous one `close to the top' with 
$\tilde{\mathcal{S}}^\dagger_A$ replaced by $\tilde{\mathcal{S}}_A$ 
and $|\Omega\rangle$ by the completely filled state.

In existing studies of the fermionic sector so far the fermions 
have been treated `quasi-classically', that is, as Grassmann valued 
$c$-numbers. However, the correspondence between the full supergravity 
equations of motion and the fermionic extension of the $E_{10}/K(E_{10})$ 
model, as far as it has been established, is lacking inasmuch as the relevant 
spinorial representations of the `R-symmetry' group $K(E_{10})$ identified 
to date are all {\em unfaithful}, and hence cannot capture the full 
fermionic dynamics of M-theory (see \cite{Damour:2006xu} for a detailed 
discussion of this problem). Again, quantisation may be essential here;
in fact, a satisfactory solution and incorporation of all the fermionic 
degrees of freedom into the $E_{10}$ model may require some kind of 
`bosonization' of the fermionic degrees of freedom. This would also be 
in accord with the fact that fermions are intrinsically  quantum objects.
\end{section}

\begin{section}{Outlook}
The cosmological billiards description takes into account
the dependence on spatial inhomogeneities and matter degrees of freedom
only in a very rudimentary way via the infinite potential walls. It would
thus be desirable to develop an approximation scheme for the quantum state
in line with the `small tension' expansion proposed in \cite{Damour:2002cu}, 
and thereby hopefully resolve the difficulties encountered in extending 
the `dictionary' of \cite{Damour:2002cu} to higher order spatial gradients 
and heights of roots in a quantum mechanical context. In the strict BKL approximation,
the full wavefunction is expected to factorize as (see also~\cite{Kirillov:1995wy})
\begin{eqnarray}\label{FullPsi}
|\Psi_{\text{full}}\rangle \sim \prod_{\bf x} |\Psi_{\bf x}\rangle\,,
\end{eqnarray}
near the singularity into a formal ultralocal product over wavefunctions of 
type (\ref{formalsol}), one for each spatial point, with independent 
bosonic wavefunctions $\Phi_{\bf{x}}\big(\rho({\bf{x}}),z({\bf{x}})\big)$ 
and {\em space-dependent} metric variables 
$\beta^a({\bf{x}})\equiv (\rho({\bf{x}}),z({\bf{x}}))$. At first sight, it may 
seem paradoxical that (\ref{FullPsi}) should become a better and better
approximation near the singularity, as all the dynamics gets concentrated
in a continuous superposition of Cartan subalgebras, whereas one would 
expect the full tower of $E_{10}$ Lie algebra states, rather than just
the Cartan subalgebra degrees of freedom, to become excited ---
in analogy with string cosmology, where one would expect the full tower of
string states to become relevant near the singularity. However, the 
apparent paradox may well turn out to be the crux  of the matter: the task 
is to replace the formal expression (\ref{FullPsi}) involving a formal 
continuous product over all spatial points by a wavefunction which solely 
depends on the (infinite) tower of $E_{10}$ degrees of freedom, and where
all spatial dependence is discarded. It is this step which would effectively 
implement the de-emergence of space and time near the cosmological 
singularity, and their replacement by purely algebraic 
concepts~\cite{Damour:2005zs,Damour:2008zza}.

For this purpose we will have to generalize the mini-superspace
Hamiltonian (\ref{WDWHam}) to the full $E_{10}$ Lie algebra. 
In fact, as a consequence of the uniqueness of the quadratic Casimir
operator on $E_{10}$, there is a {\em unique} $E_{10}$ extension of 
the billiard Hamiltonian (\ref{WDWHam}) given by
\begin{eqnarray}\label{E10Ham}
\mathcal{H}_0 \to \mathcal{H} \, = \, \mathcal{H}_0 \,  + 
\sum_{\alpha\in\Delta_+(E_{10})}\sum_{s=1}^{\text{mult}(\alpha)} 
e^{-2\alpha(\beta)}\Pi_{\alpha,s}^2\,.
\end{eqnarray}
where the first sum runs over the positive roots $\alpha$ of $E_{10}$ 
and the second one over a basis of the possibly degenerate root space 
of $\alpha$. Due to our lack of a manageable realization of the 
$E_{10}$ algebra, this is a highly formal expression, but we can
nevertheless note two important features: like the 
mini-superspace Hamiltonian (\ref{WDWHam}) this operator is free 
of ordering ambiguities by the uniqueness of the
$E_{10}$ Casimir operator, and it has the form of a {\em free} 
Klein-Gordon operator, albeit in infinitely many dimensions. The 
absence of potential terms is due to the fact that in the approach of 
\cite{Damour:2002cu} (as far as it has been worked out, at least) 
the spatial gradients, which give rise to the `potential' $\propto R^{(3)}$
in the standard WDW equation, are here replaced by {\em time derivatives 
of dual fields}. Accordingly, the contributions to the potential
are replaced by momentum-like operators $\propto \Pi^2$. It is for 
this reason that the restriction to positive norm wave functions may 
still be consistent for the full $E_{10}$ theory --- unlike for the 
standard WDW equation (but we note that so far no one has succeeded
in generalizing the mini-superspace scalar product (\ref{KGProduct}) 
to the full theory).

Nevertheless, the Hamiltonian (\ref{E10Ham}) is not the complete
story because, as in standard canonical gravity, the extended system 
requires additional constraints. For the $E_{10}$ model their complete
form is not known, but a first step towards their incorporation was 
taken in \cite{Damour:2007dt} where a correspondence was 
established at low levels between the classical canonical 
constraints of $D=11$ supergravity (in particular, the 
diffeomorphism and  Gauss constraints) on the one hand, and a set of 
constraints that can be consistently imposed on the $E_{10}/K(E_{10})$ coset 
space dynamics on the other (see \cite{DKN} for more recent results
in this direction). The fact that the latter can be cast in a 
`Sugawara-like' form quadratic in the $E_{10}$ Noether charges 
\cite{Damour:2007dt} would make them particularly amenable for the 
implementation on a quantum wavefunction. In addition, one would expect 
that $PSL_2(\mathtt{O})$ must be replaced by a much larger `modular group' 
whose action extends beyond the Cartan subalgebra degrees of freedom 
all the way into $E_{10}$, perhaps along the lines suggested in \cite{Ganor}.

As noted above, the inequality (\ref{eigenvalues}) implies that  
$\Phi(\rho,z)\to 0$ for $\rho\to\infty$, and hence the wavefunction $\Psi$ 
vanishes at the singularity, in such a way that the norm is preserved. 
Its oscillatory nature entails that it cannot be analytically extended 
beyond the singularity, a result whose implications for the question 
of singularity resolution in quantum cosmology remain to be explored. 
At least formally, these conclusions remain valid in the full theory:
the extra contribution in (\ref{E10Ham}) extending $\mathcal{H}_0$ to
the full Hamiltonian $\mathcal{H}$ are positive, as follows from the manifest 
positivity of the $E_{10}$ Casimir operator on the complement of the
Cartan subalgebra of $E_{10}$. Hence the inequality (\ref{eigenvalues})
is further strengthened and thus constitutes a lower bound also for the 
full Hamiltonian.

To put our results in perspective we recall that the mechanism 
usually invoked to resolve singularities in canonical approaches
to quantum geometrodynamics would be to replace the classical `trajectory' 
in the moduli space of 3-geometries (that is, WDW superspace) by a quantum 
mechanical wave functional which `smears' over the singular 3-geometries. 
By contrast, the present work suggests a very different picture, namely 
the `resolution' of the singularity via {\em the effective disappearance 
(de-emergence) of space-time} near the singularity (see also 
\cite{Damour:2008zza}). The singularity would 
thus become effectively `unreachable'. This behavior is very 
different from other possible mechanisms, such as the Hartle--Hawking no 
boundary proposal \cite{Hartle:1983ai}, or cosmic bounce scenarios of the type
considered recently in the context of minisuperspace loop quantum cosmology 
\cite{Bojowald:2001xe,Ashtekar:2007em,Bojowald:2008ec}, both of which
require continuing the cosmic wavepacket into and beyond the singularity
at $\rho=\infty$. In contrast to these models, the exponentially growing 
complexity of the $E_{10}$ Lie algebra suggests that it may turn out
to be impossible to `resolve' the quantum equations as one gets closer
and closer to the singularity. In other words, there may appear an element 
of {\em non-computability} (in a mathematically precise sense) that may
forever screen the big bang from a complete resolution.

A key question for singularity resolution concerns the role of observables, 
and their behavior near the singularity. While no observables (in the sense 
of Dirac) are known for canonical gravity, we here only remark that
for the $E_{10}/K(E_{10})$ coset model the conserved $E_{10}$ Noether
charges do constitute an infinite set of observables, as these charges
can be shown to commute with the full $E_{10}$ Hamiltonian (\ref{E10Ham}).
The expectation values of these charges are the only quantities that 
remain well-defined and can be sensibly computed in the deep 
quantum regime, where the $E_{10}/K(E_{10})$ coset model is expected to
replace space-time based quantum field theory. In the final analysis
these charges would thus replace geometric quantities (such as the 
curvature scalar) which blow up at the singularity, but which are 
not canonical observables.\\
\end{section}

\noindent
{\bf Acknowledgements:}  H.N. is grateful to the Institute for Advanced Studies
in Stellenbosch, South Africa, and especially D.~Loureiro, J.~Murugan and 
A.~Weltman for hospitality and for  organizing a wonderful conference 
in honor of George Ellis' 70th birthday. We thank J.~Barbour, M.~Berry, 
M.~Koehn, R.~Penrose and H.~Then for discussions and correspondence. 
AK is a Research Associate of the Fonds de la Recherche 
Scientifique--FNRS, Belgium.

\end{document}